\documentclass[aps,twocolumn,pre,showpacs,eqsecnum]{revtex4}
\usepackage{graphpap}
\usepackage[dvips]{graphicx}
\usepackage[dvips]{graphics}
\usepackage{color}

\begin{document}

\title{Energy gaps in etched graphene nanoribbons}
 \author{ C. Stampfer, J. G\"uttinger, S. Hellm\"uller, F. Molitor, K. Ensslin and T. Ihn} 
 \affiliation{Solid State Physics Laboratory, ETH Zurich, 8093 Zurich, Switzerland}
\date{ \today}

 \begin{abstract}
Transport measurements on an etched graphene nanoribbon are presented.  
It is shown that two distinct voltage scales can be experimentally extracted that characterize the parameter region of suppressed conductance at low charge density in the ribbon. One of them is related to the charging energy of localized states, the other to the strength of the disorder potential. The lever arms of gates vary by up to 30\% for different localized states which must therefore be spread in position along the ribbon. A single-electron transistor is used to prove the addition of individual electrons to the localized states.  
In our sample the characteristic charging energy is of the order of 10 meV, the characteristic
strength of the disorder potential of the order of 100 meV.
 \end{abstract}

 \maketitle

\newpage
Graphene nanoribbons~\cite{che07,han07,dai08,wan08,lin08} and narrow graphene constrictions~\cite{sta08a,pon08,sta08b}
display unique electronic properties 
based on
truly two-dimensional (2D)
graphene~\cite{gei07} with potential applications in nanoelectronics~\cite{kat07} and spintronics~\cite{tra07}. 
Quasi-1D graphene nanoribbons and constrictions are of 
interest due to the presence of an effective energy gap, overcoming
the gap-less band structure of graphene and leading to overall semiconducting
behavior, most promising for the fabrication of nanoscale graphene transistors~\cite{wan08}, tunnel barriers, and quantum dots~\cite{sta08a,sta08b,pon08}.
On the other hand, ideal graphene nanoribbons~\cite{bre06,whi07}
promise interesting quasi-1D physics with strong relations to
carbon nanotubes~\cite{rei03}. 
Zone-folding approximations~\cite{whi07}, $\pi$-orbital tight-binding models~\cite{per06,dun07},
and first principle calculations~\cite{fer07,yan07a} predict an energy gap $E_\mathrm{g}$ scaling as $E_\mathrm{g} = \alpha/W$ with the nanoribbon width $W$, 
where $\alpha$ ranges between 0.2--1.5~eV$\times$nm, depending on the model and the crystallographic orientation of the nanoribbon~\cite{lin08}.
However, these theoretical estimates 
can neither explain the experimentally observed energy gaps of
etched nanoribbons of widths beyond 20\,nm, which turn out to be larger
than predicted, nor do they explain the large number of resonances found inside the gap~\cite{che07,han07,sta08b}.
This has led to the
suggestion that localized states (and interactions effects) due to edge roughness,
bond contractions at the edges~\cite{son08} and disorder
may dominate the transport gap.
%
%
%
%
%
   \begin{figure}[t]\centering
\includegraphics[draft=false,keepaspectratio=true,clip,%
                   width=1\linewidth]%
                   {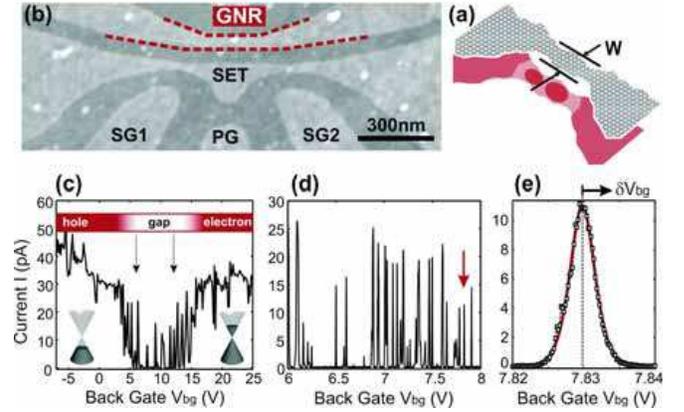}                   
\caption[FIG1]{(color online)
(a) Schematic illustration of an etched nanoribbon with width $W$, also highlighting local charge islands along the nanoribbon.
(b)  Scanning force microscope image of an etched graphene nanoribbon (GNR) with a nearby single electron transistor (SET) and lateral graphene gates (PG, SG1 and SG2).  (c) Low bias ($V_b=300 \mu V$) back gate characteristics of the GNR showing that the regimes of hole and electron transport are separated by the transport gap,  delimited by the vertical arrows. (d) High resolution close-up inside the gap displaying a large number of sharp resonances within the gap region. (e) Close-up of a single resonance (see arrow in panel (d)).} 
\label{trdansport}
\end{figure}
%
%
Several mechanisms have been proposed to describe the 
 observed gap,
including re-normalized lateral confinement~\cite{han07}, quasi-1D Anderson localization~\cite{muc08}, percolation models~\cite{ada08} and many-body effects (incl. quantum dots)~\cite{sol07}, where substantial edge disorder is required.  Recently, it has been shown that also moderate amounts of edge roughness can substantially suppress the linear conductance near the charge neutrality point~\cite{hei08}, giving rise to localized states relevant for both single particle and many-body descriptions.

In this paper
 we show experimental evidence that the transport gap
in an etched graphene nanoribbon
(see schematic in Fig.~1a) is primarily formed by local resonances and quantum dots along the ribbon. 
We employ lateral graphene gates to show that size and location of individual charged islands in the ribbon vary as a function of the Fermi energy.
 In addition, we use a graphene single electron transistor (SET) to 
detect individual charging events inside the ribbon.

We focus on
an all-graphene setup, as shown in Fig.~1b, where a nanoribbon (highlighted by dashed lines) with $W\approx45$~nm is placed at a distance of $\approx$~60~nm from a
graphene SET with an island diameter of $\approx$~200~nm. 
The back gate (BG) allows us to tune the overall Fermi level and three lateral graphene gates~\cite{mol07}, PG, SG1 and SG2 are used to locally tune the potential of the nanoribbon and the SET. 
The sample fabrication is based on mechanical exfoliation of graphite~\cite{nov04}, electron beam lithography, reactive ion etching and evaporation of Ti/Au
contacts. A detailed description is found in Refs.~\onlinecite{dav07a,mol07}, and \onlinecite{sta08a}.
%
%
%
%
The device is measured in two-terminal geometry by low frequency lock-in techniques in a variable temperature insert cryostat at a temperature of 1.7~K.

\begin{figure}[t]\centering
\includegraphics[draft=false,keepaspectratio=true,clip,%
                   width=0.9\linewidth]%
                   {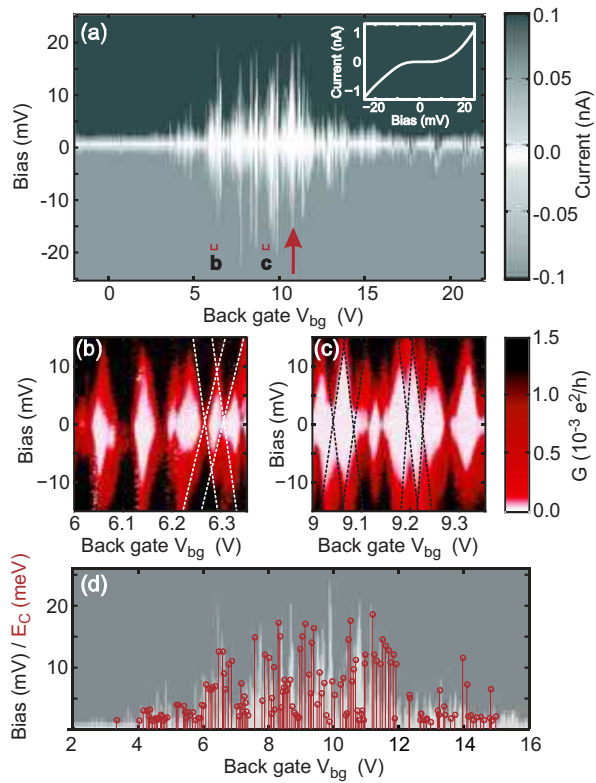}                   
\caption[FIG2]{ 
(color online) (a) Source-drain current measurements as function
of bias and back gate voltage (all other gates have been grounded) on the 45~nm wide nanoribbon (Fig.~1a).
The white areas are regions of
strongly suppressed current forming the energy gap. The inset shows a typical non-linear I-V characteristic
($V_{bg}$ = 10.63 V, see arrow). (b,c) Differential conductance ($G$) measurements as close-ups of
panel (a) at two different back gate regimes (see labels in (a)). These measurements show diamonds with suppressed conductance (highlighted by dashed lines) allowing to extract the charging
energy from individual diamonds. (d) Individual charging energies as function of the back gate voltage over a wide range plotted on top
of half of the back gate voltage range shown in panel (a). 
}
\label{expresults}
\end{figure}

Fig.~1c shows the low source--drain bias ($V_b$ = 300 $\mu$V $<\!\!< 4 k_B T$) back-gate characteristic of the nanoribbon,
where we tune transport from the hole (left side) to the electron
regime. The region 6~V $< V_{bg} <$ 12~V of suppressed current (marked by two arrows) is the so-called transport gap $\Delta V_{bg}$ in back gate voltage ($\Delta V_{bg} \approx$ 6~V).
In contrast to the energy gap predicted for samples without disorder, where transport should be completely pinched-off, we observe---in good agreement with other experimental work~\cite{che07,han07,sta08b}---a large number of reproducible conductance resonances
inside the gap.

A high-resolution close-up of Fig.~1c shown in Fig.~1d reveals a sequence of resonances with a small line-width indicating strong localization.
A particularly narrow resonance is shown in Fig.~1e (see arrow in Fig.~1d). 
The line-shape can be well fitted by $I \propto \cosh^{-2}(e \alpha_{bg} \delta V_{bg} /2.5 k_B T_e)$,  
 where $\alpha_{bg} \approx 0.2$ is the back gate lever arm and $\delta V_{bg} = V_{bg}-V_{bg}^{peak}$ (see Fig.~1e)~\cite{bee91}.
The estimated effective electron temperature, $ T_e~\approx$~2.1~K, is close to the base temperature, leading to the conclusion
that the peak broadening is mainly limited by temperature rather than by the life-time of the
resonance.

In Fig.~2a we show source--drain current measurements on the nanoribbon as a function of source--drain bias and back gate voltage (i.e., Fermi energy). We observe regions of strongly suppressed current (white areas) leading to an effective energy gap in bias direction inside the transport gap in back gate voltage (shown in Fig.~1c). Highly non-linear I-V characteristics (see e.g. inset in Fig.~2a) are characteristic for the energy gap in bias direction. 
This energy gap agrees reasonably well with the observations
in Refs.~\onlinecite{han07}, and \onlinecite{sol07} of 
an 
 energy gap
of $E_g \approx$ 8~meV, for $W$ = 45 nm.

The transport gap in source--drain bias voltage corresponding to the energy gap $E_g$, and the transport gap $\Delta V_{bg}$ in back gate voltage are two distinct voltage scales resulting from our experiment. The quantity $\Delta V_{bg}$ is measured at constant (nearly zero) $V_b$ (transport window) but varying Fermi energy $E_F$ and is related to a change in Fermi energy $\Delta E_F$ in the system. Varying the magnitude of the transport window $V_b$ at fixed Fermi energy gives rise to $E_g$.

We estimate the energy scale $\Delta E_F$ corresponding to $\Delta V_{bg}$ from
$\Delta E_F \approx \hbar v_F \sqrt{2\pi C_g \Delta V_{bg}/\left|e\right|}$,
 (where $C_g$ is the
back gate capacitance per area)~\cite{com01a}.
We find an energy gap $\Delta E_F \approx 110-340$~meV which is more than one order of magnitude larger than $E_g$.
We attribute this discrepancy 
to different physical meanings of these two energy scales, which will be further illustrated and discussed below.

More insight into the two energy scales and their relation is gained
by focusing on 
a smaller back gate voltage range
as shown
in Figs.~2b,c which are two high resolution differential conductance $dI/dV_b$ close-ups of Fig.~2a (see labels therein).
 At this scale transport is dominated by well distinguishable diamonds of suppressed conductance (see bright areas and dashed lines in Figs.~2b,c) which indicate 
 that transport is blocked by localized electronic states
or quantum dots (see also Ref.~\cite{mol08a}). The related charging energy $E_c$ which itself is related to the quantum dot size, depends on the Fermi energy  on a small back gate voltage scale (see different diamond sizes in Figs.~2b,c), but also on a large scale (see Fig.~2a). In order to confirm this statement
 %
the extracted charging energies are plotted in Fig.~2d  into the top half of the measurements shown in Fig.~2a.

\begin{figure}[t]\centering
\includegraphics[draft=false,keepaspectratio=true,clip,%
                   width=0.98\linewidth]%
                   {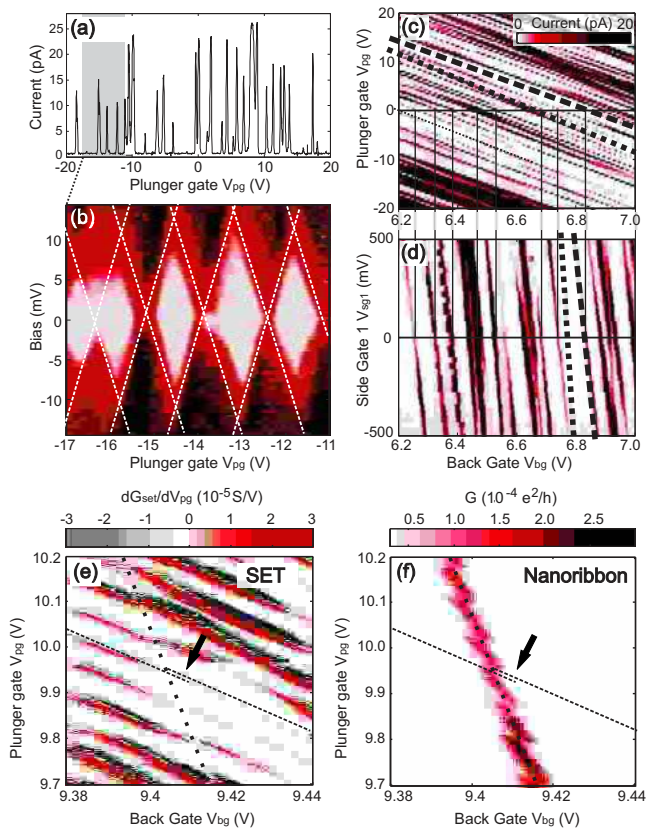}                   
\caption[FIG2]{
(color online) (a) Low bias ($V_b=300 \mu V$) current measurements as function of plunger gate voltage at fixed back gate ($V_{bg}$ =~7 V), showing a large number of sharp resonances within the gap region. 
(b) Corresponding diamonds (see highlighted area in panel (a)) in differential conductance $G$ (same color scale as in Figs.~2b,c, bright regions represent low conductance). Here, a dc bias $V_b$ with a small ac modulation (50~$\mu$V) is applied symmetrically across the nanoribbon. (c,d) Charge stability diagrams as function of plunger gate and back gate voltage (c) and side gate 1 and back gate voltage (d). These plots highlight that individual resonances have individual lever arms (see dashed and dotted lines).
(e,f) Detection of individual charging events in the nanoribbon by the nearby single electron transistor (SET). (e) Coulomb blockade resonances on the SET as function of $V_{pg}$ and $V_{bg}$ exhibit clear signatures of the charging event in the nanoribbon expressed by crossing the local resonance (f). 
Dotted and dashed lines show the different lever arms. 
}
\label{expresults}
\end{figure}

Figures.~3a,b show differential conductance measurements at fixed
$V_{bg}=7$~V, as a function of
the lateral plunger gate voltage (c.f., Fig.~1b) which tunes the potential on the nanoribbon locally.
%
Similar to the back gate dependent measurements in Fig.~1d we observe  in Fig.~3a a large number of resonances inside the transport
gap. In contrast to back gate sweeps we find certain plunger gate regions with
almost equally spaced conductance peaks (see, e.g., the highlighted regime in Fig.~3a and the corresponding diamonds in Fig.~3b),
giving rise to the assumption that here only a single charged island is tuned by the lateral gate. These diamond measurements are of comparable quality as those presented in Refs.~\cite{sta08a,pon08,sta08b}.

By following resonances at low bias ($V_{b}=500 \mu V$) over
a larger $V_{pg}-V_{bg}$ range (see Fig.~3c) we observe that individual resonances exhibit different relative lever arms in the range of $\alpha_{pg,bg} \approx 0.039-0.048$ (slopes of dotted and dashed lines in Fig.~3c). 
These variations of up to 20$\%$ can be attributed to different 
capacitances between the plunger gate and individual electron puddles,
which sensitively includes their local position on the ribbon. By sweeping the voltage on the more asymmetrically placed side gate~1 (see Fig.~1b) rather than the plunger gate  this effect is even enhanced. In Fig.~3d we show the corresponding measurements ($V_{pg}$ = 0 V). Relative lever arms in the range of $\alpha_{sg1,bg} \approx 0.054-0.077$ with scattering of more than 30$\%$ can be extracted.
The stability of the
sample allows to match resonances seen in Figs.~3c and 3d so that they can be followed in a 3D parameter space. These measurements confirm local resonances being located along the nanoribbon, with different lever arms to the local lateral gates.    
%

We now make use of the SET device fabricated near the ribbon to detect individual charging events inside
localized states of the nanoribbon. The SET which has been characterized before~\cite{gue08}, has a charging energy of $E_{c,SET} \approx 4.5$~meV and Coulomb blockade peak spacing fluctuations below $15\%$. The Coulomb resonances in the conductance of the SET,
highlighted as dashed lines in Fig.~3e can be used to detect charging of a local resonance (dotted line in Fig.~3e) in the nanoribbon with individual electrons. 
We show conductance measurements as function of
plunger and back gate voltage in order to identify resonances of the SET and the nanoribbon via their different relative lever arms (Figs.~3e,f).
Since the SET is much closer to the PG than the nanoribbon, the relative lever arm $\alpha_{pg,SET}/\alpha_{bg,SET} \approx$~0.18 seen as the slope of SET resonances in Fig.~3e (dashed lines in Figs. 3e,f) is significantly
larger than the relative lever arm of a resonance in the nanoribbon $\alpha_{pg,bg} \approx$~0.04 shown in Fig. 3f (dotted lines in Figs. 3e,f).
The observation of a jump (see arrow in Fig.~3e) in the Coulomb resonances of the SET when they cross the resonance in the ribbon
is a clear signature of charging the localized state in the nanoribbon, which  changes in a discontinuous way the potential on the SET island by capacitive coupling. This shows that we accumulate localized charges along the nanoribbon as function of the back gate voltage. 

The experimental data shown above provide strong indications that the two experimentally observed energy scales $E_g$ and $\Delta E_F$ are related to charged islands or quantum dots forming spontaneously along the nanoribbon. This is supported by the observation (i) of Coulomb diamonds, which vary in size as function of the Fermi energy, (ii) of a strong variation of the relative lever arms of individual resonances and (iii) of local charging of islands inside the nanoribbon.

Quantum dots along the nanoribbon can arise in the presence of a quantum
confinement energy gap ($\Delta E_{con}$) combined with a strong bulk and edge-induced disorder potential $\Delta_{dis}$, as illustrated in Fig.~4.
The confinement energy can be estimated by $\Delta E_{con} (W) \approx \gamma \pi a_{C-C}/W$, where $\gamma \approx 2.7$~eV and $a_{C-C}=0.142$~nm~\cite{whi07}. This leads to $\Delta E_{con} =26$~meV for $W=45$~nm, which by itself can neither explain the observed energy scale $\Delta E_F$, nor the formation of quantum dots in the nanoribbon.
However, by superimposing a disorder potential giving rise to electron-hole puddles near the charge neutrality point~\cite{mar07}, the confinement gap ensures that Klein tunneling (from puddle to puddle) gets substituted by real tunneling.
Within this model $\Delta E_F$ depends on both
the confinement energy gap and the disorder potential.
An upper bound for the magnitude of the disorder potential can be estimated from our data to be given by $\Delta E_F$.
Comparing to Ref.~\cite{mar07} where a bulk carrier density fluctuation of the order of $\Delta n \approx \pm 2 \times  10^{11}$~cm$^{-2}$ was reported, we find reasonable agreement as the corresponding variation of the local potential is $\Delta E_F \approx 126$~meV.

We can estimate the fraction of overlapping diamonds by summing over all charging energies $E_c$
observed in Fig.~2d. This leads to $\sum E_c \approx$~630~meV. Comparison with the estimate for $\Delta E_F$ gives 45 - 82 $\%$ overlapping diamonds. We expect that this value depends strongly on the length of the nanoribbon
in agreement with findings of Ref.~\cite{mol08a}.

The energy gap in bias direction $E_g$ does not tell much about the magnitude of the disorder potential, but it is rather related to the sizes of the charged islands. In particular, the minimum island size is related to the maximum charging energy $E_{c,max}$. By using a simple disc model we can estimate the effective charge island diameter by $d = e^2/(4 \epsilon \epsilon_r E_c) \approx 100$~nm (where $\epsilon =(1+4)/2$), which exceeds the nanoribbon width $W$. Thus, in ribbons of different width the charging energy will scale with $W$ giving the experimentally observed $1/W$ dependence of the energy gap in bias direction
~\cite{han07}.


\begin{figure}[t]\centering
\includegraphics[draft=false,keepaspectratio=true,clip,%
                   width=1.0\linewidth]%
                  {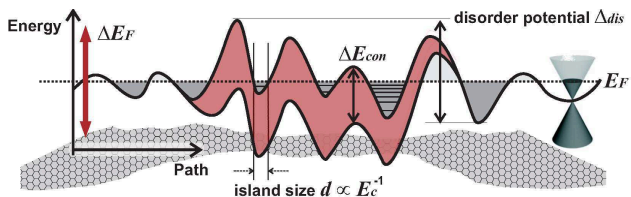}                   
\caption[FIG2]{
(color online) Schematic illustration of the potential landscape along the graphene
nanoribbon allowing the formation of charged islands and quantum dots. For more information
see text.  
}
\label{expresults}
\end{figure}

In conclusion, we have presented detailed transport
measurements on a graphene nanoribbon, focusing on
the origin of the transport gap.
Experimentally we find two distinct energy scales characterizing this gap.
The first of them is the charging energy of local resonances or
quantum dots forming along the ribbon. The second is probably dominated by the strength of
the disorder potential, but also dependent on the gap induced by confinement due to the ribbon boundaries. 
These insights are important to understand transport in graphene nanostructures and
may help in designing future graphene nanoelectronic components.

{Acknowledgment ---}
The authors wish to thank A. Castro-Neto, S. Das Sarma, T. Heinzel, M.~Hilke, F.~Libisch,
K.~Todd and L. Vandersypen  
 for helpful discussions. Support by the Swiss
National Science Foundation and NCCR nanoscience are gratefully 
acknowledged.


\begin{thebibliography}{99}

\bibitem{che07}
Z. Chen, Y.-M. Lin, M. Rooks and P. Avouris, Physica E,~{\bf 40}, 228, (2007).

\bibitem{han07}
M. Y. Han, B. \"Ozyilmaz, Y. Zhang, and P. Kim, Phys. Rev. Lett.,~{\bf 98}, 206805 (2007)

\bibitem{dai08}
X. Li, X. Wang, L. Zhang, S. Lee, H. Dai, Science,~{\bf 319}, 1229 (2008).

\bibitem{lin08}
Y.-M. Lin, V. Perebeinos, Z. Chen and P. Avouris, arXiv:080xx0035v2 (2008)

\bibitem{wan08}
X. Wang, Y. Ouyang, X. Li, H. Wang, J. Guo, and H. Dai, Phys. Rev. Lett.,~{\bf 100}, 206803 (2008)

\bibitem{sta08a}
C. Stampfer, J. G\"uttinger, F. Molitor, D. Graf, T. Ihn, and K. Ensslin, Appl. Phys. Lett.,~{\bf 92}, 012102 (2008)

\bibitem{pon08}
L. A. Ponomarenko, F. Schedin, M. I. Katsnelson, R.~Yang, E.~H.~Hill, K.~S.~Novoselov, A.~K.~Geim,
Science,~{\bf 320}, 356 (2008).

\bibitem{sta08b}
C. Stampfer, E. Schurtenberger, F. Molitor, J. G\"uttinger, T. Ihn, and K. Ensslin, Nano Lett.,~{\bf 8}, 2378 (2008)

\bibitem{gei07}
A.~K. Geim and K.~S. Novoselov, Nat. Mater. {\bf 6}, 183
(2007).

\bibitem{kat07}
M.~I. Katnelson, Materials Today {\bf 10(1-2)}, 20 (2007)

\bibitem{tra07}
B. Trauzettel, D.V. Bulaev, D.~Loss, and G.~Burkard, Nature Physics,~{\bf 3}, 192, (2007).



\bibitem{bre06}
L. Brey and H. A. Fertig, Phys. Rev. B,~{\bf 73}, 235411 (2006).


\bibitem{whi07}
C. T. White, J. Li, D. Gunlycke, and J. W. Mintmire, Nano Lett.,~{\bf 7}, 825 (2007). 

\bibitem{rei03} For review on carbon nanotubes see e.g.:
S. Reich, C. Thomsen, J. Maultzsch, "Carbon Nanotubes", Wiley-VCH, 2003.


\bibitem{per06}
N. M. R. Peres, A. H. Casrtro Neto and F.~Guinea, Phys. Rev. B,~{\bf 73}, 195411 (2006).

\bibitem{dun07}
D. Dunlycke, D. A. Areshkin, and C. T. White,  Appl. Phys. Lett.,~{\bf 90}, 142104 (2007).

\bibitem{fer07} 
J.~Fernandez-Rossier, J~.J.~Palacios, and L.~Brey, Phys. Rev. B,~{\bf 75}, 205441 (2007).

\bibitem{yan07a}
L.~Yang, C.-H.~Park, Y.-W. Son, M.~L.~Cohen, and S.~G.~Louie, Phys. Rev. Lett.,~{\bf 99}, 186801 (2007).

\bibitem{sol07}
F.~Sols, F.~Guinea and A.~H.~Castro Neto, Phys. Rev. Lett.,~{\bf 99}, 166803 (2007).

\bibitem{son08}
Y.-W. Son, M.~L.~Cohen, and S.~G.~Louie, Phys. Rev. Lett.,~{\bf 99}, 186801 (2007).

\bibitem{muc08}
E.~R. Mucciolo, A.~H.~Castro Neto and C.~H.~Lewenkopf, arXiv:0806.3777v1 (2008).

\bibitem{ada08}
S. Adam and S. Cho and M. S. Fuhrer and S. Das Sarma,
Phys. Rev. Lett.,~{\bf 101}, 046404 (2008).

\bibitem{hei08}
M. Evaldsson, I. V. Zozoulenko, Hengyi Xu, T. Heinzel, arXiv:0805.4326 (2008).

\bibitem{mol07}
F. Molitor, J. G\"uttinger, C. Stampfer, D. Graf, T. Ihn, and
K. Ensslin,  Phys. Rev. B {\bf 76}, 245426 (2007).

\bibitem{nov04}
K. S. Novoselov, A. K. Geim, S. V. Morozov, D.~Jiang, M.~I.~Katsnelson, S.~V.~Dubonos, 
I.~V.~Grigorieva, A. A. Firsov, Science,~{\bf 306}, 666, (2004).

\bibitem{dav07a} 
D. Graf, F. Molitor, K. Ensslin, C. Stampfer, A. Jungen, C. Hierold, and
L. Wirtz, Nano Lett. {\bf 7}, 238 (2007).


\bibitem{bee91}
C. W. J. Beenakker, Phys. Rev. B {\bf 44}, 1646 (1991).

\bibitem{mol08a} 
F.~Molitor et al., in preparation (2008).

\bibitem{com01a}
Here we assume that the charge neutrality point is right in the center of the
transport gap ($V_{bg}^{D} \approx $9~V). Moreover, we make use of the following
boundary conditions for $C_g(W)$. Lower bound: $C_g(W\rightarrow\infty)=C_{g,2D}\approx 7.2 \times 10^{10}$ cm$^{-2}$V$^{-1} \left|e\right|$. Upper bound: Following Ref.~\cite{lin08} the ratio $C_g(W)/C_{g,2D}$ increases with decreasing $W/d$, where $d$ is the oxide thickness, and we obtain $C_g(W$~=~30~nm$)~\approx$ 10~$C_{g,2D}$.



\bibitem{gue08} 
J. G\"uttinger, C. Stampfer, S. Hellm\"uller, F. Molitor, T. Ihn, and
K. Ensslin, arXiv:0809.3904 (2008)

\bibitem{mar07}
J. Martin, N. Akerman, G. Ulbricht, T. Lohmann, J.H. Smet, K. von Klitzing, 
and A. Yacoby, Nature Physics,~{\bf 4}, 144 - 148 (2008).





\end{thebibliography}
\end{document}